\documentclass[twocolumn, pra, showpacs,superscriptaddress]{revtex4}
\usepackage{graphicx}
\usepackage{dcolumn}
\usepackage{bm}
\usepackage{amsmath}

\begin{document}

\title{Ground-state properties of few-Boson system in a one-dimensional hard
wall potential with split}
\author{Xiangguo Yin}
\affiliation{Department of Physics and Institute of Theoretical Physics, Shanxi
University, Taiyuan 030006, P. R. China}
\author{Yajiang Hao}
\affiliation{Institute of Physics, Chinese Academy of Sciences, Beijing 100080, P. R.
China}
\author{Shu Chen}
\affiliation{Institute of Physics, Chinese Academy of Sciences, Beijing 100080, P. R.
China}
\author{Yunbo Zhang}
\email{ybzhang@sxu.edu.cn}
\affiliation{Department of Physics and Institute of Theoretical Physics, Shanxi
University, Taiyuan 030006, P. R. China}

\begin{abstract}
We carry out a detailed examination of the ground state property of
few-boson system in a one-dimensional hard wall potential with a $\delta -$%
split in the center. In the Tonks-Girardeau limit with infinite repulsion
between particles, we use the Bose-Fermi mapping to construct the exact $N-$%
particle ground state wavefunction which allows us to study the correlation
properties accurately. For the general case with finite inter-particle
interaction, the exact diagonalization method is exploited to study the
ground-state density distribution, occupation number distribution, and
momentum distribution for variable interaction strengths and barrier
heights. The secondary peaks in the momentum distribution reveal the
interference between particles on the two sides of the split, which is more
prominent for large barrier strength and small interaction strength.
\end{abstract}

\date[Date text]{date}
\maketitle

\section{Introduction}

With the development of laser cooling and optical trapping technology, quasi
one-dimensional (1D)\ cold atom systems have been realized by tightly
confining the particle motion in two directions to zero point oscillation
\cite{Gorlitz,Moritz,Paredes,Kinoshita,Tolra,Olshanii}. Furthermore, optical
dipole forces generated by several crossed, interfering laser beams make it
possible to create periodic potentials, and experimentalists are able to
trap and control small numbers of particles in optical lattice or double
well potential. Meanwhile with the Feshbach resonance technique, the
inter-particle scattering length can be easily changed by tuning a magnetic
field, which allows the atoms to enter the full regime of interaction from
weakly to strongly interacting limit. In particular, a quasi-1D quantum gas
of strongly interacting bosons has been observed in the so-called
Tonks-Girardeau (TG) regime \cite{Paredes,Kinoshita}. These experiment
progresses inspire great interest in exploring the properties of bosonic
gases in the full interaction regime.

As is well known, the Gross-Pitaviskii (GP) theory is widely adopted in
dealing with the system of Bose-Einstein condensations with weak
interaction. In spite of its great success in explaining the basic
experimental observations, the GP equation is fundamentally based on a
mean-field approximation and fails in the strongly correlated systems \cite%
{Kolomeisky,Chen}. For arbitrary interaction from weakly to strongly
interacting limit, there exist several theoretical treatments including
analytic method, e.g. Bethe ansatz \cite{Lieb,Hao,Yang,Yang2,Hao2}, and
numerical simulation e.g. the exact diagonalization \cite%
{Deuretzbacher,Hao08}, the multi-configuration time-dependent Hartree
(MCTDH) method \cite{Cederbaum,Zollner}, discrete variable representation
(DVR) \cite{Murphy}, and so on. Bethe ansatz provides the exact solution to
some many-body systems including Bose gas \cite{Lieb,Hao}, Fermi gas \cite%
{Yang, Hao2}, Bose-Fermi mixture \cite{Yang2} as well as multi-component
cases. In the past several years, though it has made great development in
cooperation with other theories and some numerical methods, this analytical
method mainly solves the uniform system. On the other hand, numerical
techniques provide more information about the many-body system, for example,
MCTDH method can not only reveal the ground state property, but also tackle
its dynamics \cite{Zollner08}. For finite particle interactions no
analytical solution to the two-particle problem is known for a finite split
barrier in the center. DVR has been used to achieve the discretization of
the spatial coordinates for two particles, however, the generalization to
more particles are not straightforward \cite{Murphy}.

While most theoretical works assume large particle numbers, strongly
correlated atomic systems realized in current experiments are limited to
systems with a small number of particles \cite{Paredes,Kinoshita}. The exact
diagonalization method becomes more and more versatile and powerful in
dealing with the ground state problem of such \textquotedblleft
few-body\textquotedblright\ system. In this paper, we study the ground state
properties of a few spinless bosons which are confined in a 1D hard-wall
trap and interact via repulsive contact potential. In addition we add a $%
\delta $-split in the center of the trap potential which may be viewed as a
generic model for double well structure or, alternatively, as a good
approximation to the problem of a trap with an impurity at the center. While
the the optical box trap has been experimentally achieved \cite{Raizen}, the
$\delta $-split potential may be realized by adding an additional laser in
the center of the box trap. By changing the strength of the laser, one can
tune the height of the split barrier effectively. In order to get some exact
results in the strongly coupling limit, we first investigate the system by
Fermi-Bose mapping \cite{Girardeau,Girardeau2} in the infinite repulsive
limit in which the interactions between bosonic particles dominate the
physics of the system. Then we use the exact diagonalization method to deal
with the situation in the full interaction regime. In particular, the
reduced single particle density, momentum distribution and occupation of the
natural orbits are illustrated by varying the parameters such as the number
of particles, the strength of interaction between particles and the height
of split barrier.

This paper is organized as follows. In Sec.II we describe the Hamiltonian of
the few-boson system and review the single particle eigenstates. In
Sec.III,\ the TG gas is studied and we will see how the properties of the
ground state in this extreme limit reveal the implication of this few-boson
system. Subsequently Sec.IV is devoted to the exact diaganolization method
to deal with the variable interaction case. Finally we summarize our result
in Sec.V.

\section{Model Hamiltonian and Single-Particle Eigenstates}

We consider $N$ bosonic atoms of mass $m$ confined in a highly elongated
hard wall trap which can be treated as quasi-one dimensional. For the case
study of localization of particles, we split the trap by inserting a tunable
zero-ranged barrier at the trap center \cite{Murphy}. The atoms interact via
a delta potential and the Hamiltonian for this system is ($\hbar =m=1$)
\begin{equation}
H=\sum\limits_{i=1}^{N}h_{i}+\sum\limits_{i,j=1(i\neq j)}^{N}g\delta \left(
x_{i}-x_{j}\right) ,  \label{hamltion1}
\end{equation}
where the single-particle Hamiltonian $h_{i}$ is given by%
\begin{equation}
h_{i}=-\frac{1}{2}\frac{\partial ^{2}}{\partial x_{i}^{2}}+V\left(
x_{i}\right) +\kappa \delta \left( x_{i}\right) .  \label{singleparticlee}
\end{equation}
Here $V\left( x\right) $ describes a hard wall trap which is zero in the
region $\left( -a,+a\right) $ and infinite outside. The last term in the
single-particle Hamiltonian represents a $\delta $-type barrier located at
the origin and we denote the strength of this barrier by a positive
parameter $\kappa $. Another essential parameter to characterize the system
is the one-dimensional interaction strength $g$ which is determined by the
modified $s$-wave scattering length due to the strong transverse confinement
\cite{Olshanii}. The single-particle Hamiltonian can be exactly solved and
the solution can be found in standard textbook and we review the result here
again for later computation.

The eigen equation of single-particle Hamiltonian is%
\begin{equation}
\left[ -\frac{1}{2}\frac{\partial ^{2}}{\partial x^{2}}+V\left( x\right)
+\kappa \delta \left( x\right) \right] \phi \left( x\right) =E\phi \left(
x\right) .  \label{eigeneq}
\end{equation}%
The eigenfunctions are either symmetric or antisymmetric. The analytic
symmetric eigenfunctions are%
\begin{equation}
\phi _{n}\left( x\right) =\left\{
\begin{array}{l}
C [ \cos \left( px\right) -\frac{\kappa }{p}\sin \left( px\right) ] ,-a\leq
x\leq 0 \\
C [ \cos \left( px\right) +\frac{\kappa }{p}\sin \left( px\right) ] ,0\leq
x\leq a%
\end{array}%
\right.  \label{wavef1}
\end{equation}%
Here, $C$ is the normalization constant and $p$ is the wave vector of
particle determined by $p/\kappa +\tan \left( pa\right) =0$. The
corresponding energy is $E=p^{2}/2$. By contrast, the antisymmetric
eigenfunctions are unaffected by the barrier and vanish at the origin. They
are the same as the odd eigenstate of the hard wall trap without split ($%
\kappa =0$)%
\begin{equation}
\phi _{n}\left( x\right) =\frac{1}{\sqrt{a}}\sin \left[ \frac{\left(
n+1\right) \pi x}{2a}\right] ,\text{ }n=1,3,5\cdots  \label{wavef2}
\end{equation}%
The corresponding energies is determined by $E_{n}=\left( (n+1)\pi
/2a\right) ^{2}/2$. In the limit $\kappa \rightarrow 0$, energies for
symmetric states in unit of $\pi ^{2}/2a^{2}$ are $E_{n}=\left( 1/2\right)
^{2},\left( 3/2\right) ^{2},\left( 5/2\right) ^{2},\cdots $ , and those for
the antisymmetric eigenstates are $E_{n}=1,2^{2},3^{2},\cdots $ . In the
opposite limit, i.e. $\kappa \rightarrow \infty $, the former energy series
converge towards $E_{n}=1,2^{2},3^{2},\cdots $ , and each symmetric
eigenstate becomes degenerate with the next highest-lying antisymmetric
state.

\section{Tonks-Girardeau Gas}

The infinite interaction between the atoms can be represented as a
constraint on the allowed bosonic wavefunction $\Psi (x_{1},\cdots ,x_{N})$.%
\begin{equation}
\Psi (x_{1},\cdots ,x_{N})=0\text{ \ \ if }x_{i}=x_{j},1\leq i<j\leq N,
\label{contrain}
\end{equation}%
which suggests that TG gas may be viewed as a group of \textquotedblleft
free\textquotedblright\ bosons governed by the Hamiltonian $%
\sum\limits_{i=1}^{N}h_{i}$, while its wavefunction is subjected to the
constraint in Eq. (\ref{contrain}). One can see immediately that this
constraint is equivalent to the exclusive principle for a Fermi system with
antisymmetric wavefunctions. Based on this observation Girardeau \cite%
{Girardeau} found the Bose-Fermi mapping theorem that allows one to
construct wavefunction for a TG gas from the wavefunction of a free Fermi
system. In particular, the wavefunction of the bosonic system $\Psi
_{B}(x_{1},\cdots ,x_{N})$ is related to the non-interacting fermionic
wavefunction $\Psi _{F}(x_{1},\cdots ,x_{N})$ by the mapping
\begin{equation}
\Psi _{B}(x_{1},\cdots ,x_{N})=A(x_{1},\cdots ,x_{N})\Psi _{F}(x_{1},\cdots
,x_{N}).  \label{mapping}
\end{equation}%
The fermionic wavefunction can be constructed from Slater determinant $\Psi
_{F}(x_{1},\cdots ,x_{N})=\sqrt{1/N!}\det_{n=0,j=1}^{N-1,N}\left( \phi
_{n}\left( x_{j}\right) \right) $ where $\phi _{n}\left( x_{j}\right) $ are
the single particle wavefunctions obtained above. The antisymmetric function
$A(x_{1},\cdots ,x_{N})$ is defined as $A(x_{1},\cdots
,x_{N})=\prod\limits_{i>j}sgn\left( x_{i}-x_{j}\right) $ with $sgn$ the sign
function. The TG gas is exactly solvable in this way and one easily see that
the single-particle density or the thermodynamic properties are identical
for these two systems.

However, quantum correlations considerably differ from each other as can be
manifested in the reduced single-particle density (RSPD) or the momentum
distribution. Although the exact many-body wave function describing TG gas
can be written in compact form, it is a difficult task to calculate the RSPD
defined as%
\begin{eqnarray}
&&\rho \left( x,y\right) =N\int dx_{2}\cdots dx_{N}  \notag \\
&&\times \Psi _{B}^{\ast }(x,x_{2},\cdots ,x_{N})\Psi _{B}(y,x_{2},\cdots
,x_{N})
\end{eqnarray}%
and the momentum distribution
\begin{equation}
n(k)=\frac{1}{2\pi N}\int_{-\infty }^{+\infty }dx\int_{-\infty }^{+\infty
}dy\rho \left( x,y\right) \exp \left( -ik\left( x-y\right) \right)
\label{momentum}
\end{equation}%
which is normalized to one. To resolve the difficulty of multi-integration,
Pezer and Buljan \cite{Pezer} rewrote the RSPD in terms of the single
particle basis%
\begin{equation}
\rho \left( x,y\right) =\sum\limits_{i,j=1}^{N}\phi _{i}^{\ast }\left(
x\right) B_{ij}\left( x,y\right) \phi _{j}\left( y\right) .  \label{density2}
\end{equation}%
The coefficients $B_{ij}\left( x,y\right) $ constitute an $N\times N$ matrix
$\mathbf{B}\left( x,y\right) =\left( \mathbf{P}^{-1}\right) ^{T}\det \mathbf{%
P}$ and the entries of the matrix $\mathbf{P}$ are $P_{ij}\left( x,y\right)
=\delta _{ij}-2\int_{x}^{y}dx^{\prime }\phi _{i}^{\ast }\left( x^{\prime
}\right) \phi _{j}\left( x^{\prime }\right) $, assuming $x<y$ without loss
of generality. This scheme enables efficient and exact numerical calculation
of quantum correlations in few body systems.

\begin{figure}[tbp]
\includegraphics[width=3.0in]{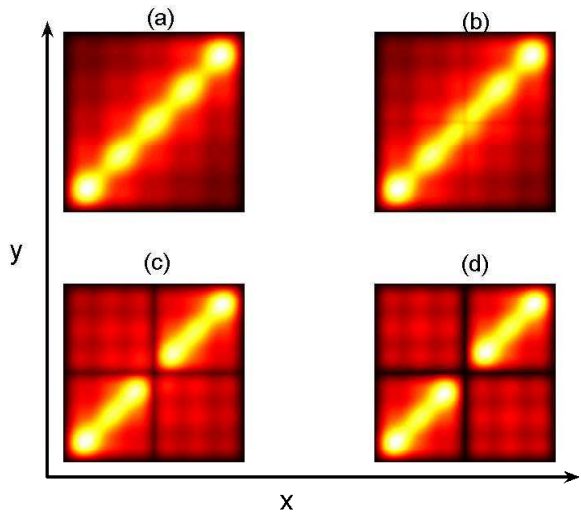}
\caption{(Color online) Reduced single-particle density matrix of 5 bosons
in TG limt, for barrier strength (a)$\protect\kappa =0.2$, (b)$\protect%
\kappa =2$, (c)$\protect\kappa =20$, (d)$\protect\kappa =200$. Each plot
spans the range $-a<x,y<a$.}
\label{fig1}
\end{figure}

\begin{figure}[tbp]
\includegraphics[width=3.0in]{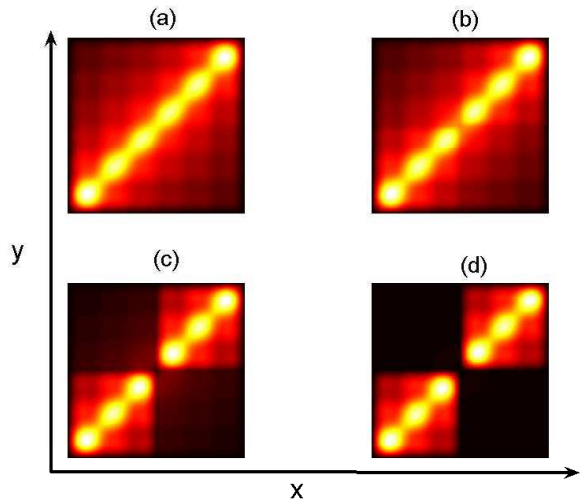}
\caption{(Color online) Reduced single-particle density matrix of 6 bosons
in TG limt, for barrier strength (a)$\protect\kappa =0.2$, (b)$\protect%
\kappa =2$, (c)$\protect\kappa =20$, (d)$\protect\kappa =200$. Each plot
spans the range $-a<x,y<a$.}
\label{fig2}
\end{figure}

We illustrate the RSPD of 5 particles in Fig. 1 and that of 6 particles in
Fig. 2 for four different barrier strengths $\kappa =0.2,2,20,200$,
respectively. The diagonal term $\rho \left( x,x\right) $ is nothing but the
single-particle density and is normalized to $N$. For a negligibly small
split barrier, Fig.1 (a) and Fig.2 (a) show that the diagonal line have $N$
peaks meaning $N$ density maxima, since the infinite interaction between
particles repel each other to occupy different positions. Meanwhile the
off-diagonal elements are non-zero, which reflect the correlation between
particles. Physically this self correlation $\rho \left( x,y\right) $ can be
viewed as the probability that two successive measurements will find the
particles at position $x$ and $y$, respectively. Increasing the barrier
strength, as seen in Fig.1 (b-d) and Fig.2 (b-d), leads to the emergence of
a quadrant separation which coincides with the result of Ref. \cite{Murphy}.
We note also in this system there appears an interesting parity effect for
odd and even number of particles. When $N$ is odd, the middle peak begins to
split into two with increasing barrier strength, so the number of density
peaks in each well increases from $\left( N-1\right) /2$ to $\left(
N+1\right) /2$. When $N$ is even, on the other hand, there are always $N/2$
peaks in each well for any value of barrier height. For a very strong split
in the center, e.g. $\kappa =200$, the off-diagonal quadrants diminish for $%
N $ is even, while remain nonzero for $N$ is odd. This implies that for odd
number of particles, the correlation between the two sides of the split
persists for a rather high barrier.

We can explain our result in a way similar to the energy band theory. The
barrier in our system splits the hard wall trap into a double-well
structure. Just as the extension of two-site Hubbard model to a lattice
gives naturally the energy band, one can imagine that the hard wall trap
with periodically distributed $\delta -$splits, i.e., Kronig-Penny
potential, admits also similar band structure \cite{Lin}. Each band in our
system, however, consists of only two energy levels. The infinite
interaction between particles forbids particles to occupy the same level and
only two particles are allowed to populate in each energy band. For the
ground state, particles occupy the energy levels from the bottom one by one.
When $N$ is even, all occupied energy bands are full and the system is in an
insulator state, while for $N$ is odd, the highest occupied band is half
filled which reminds us the conduction band in metal conductor. Particles
occupying that band can move more freely which is the reason for\ different
off-diagonal elements of RSPD induced by the parity of particle number.

These results can also be understood in a simpler way. When $N$ is even,
each well can keep $N/2$ particles, and the chemical potential can keep
balance in two sides of middle barrier. When $N$ is odd, supposed\ each well
is firstly filled with $\left( N-1\right) /2$ particles, one would face a
dilemma on how to put the last one. The probability of finding the last
particle is the same in each well, which keeps the connection between the
two wells and leads to non-zero RSPD in off-diagonal quadrant.

To characterize some BEC-like coherence effects in few body system, one
usually examine the eigenvalues and eigenfunctions of the RSPD \cite%
{Girardeau2}. The eigenfunctions of the RSPD, called the natural orbits $%
\varphi \left( x\right) $ in TG systems
\begin{equation}
\int dy\rho \left( x,y\right) \varphi _{i}\left( y\right) =\lambda
_{i}\varphi _{i}\left( x\right) ,i=0,1,2\cdots   \label{nb}
\end{equation}%
represent a set of effective single-particle states, while eigenvalues $%
\lambda _{i}$ represent their occupation and $\sum_{i}\lambda _{i}=N$. The
momentum distribution is related to these natural orbits via the relation $%
n(k)=\sum_{i}\lambda _{i}|\mu _{i}\left( k\right) |^{2}$, where $\mu
_{i}\left( k\right) $ denotes the Fourier transformation of $\varphi
_{i}\left( x\right) $
\begin{equation*}
\mu _{i}\left( k\right) =\int dx\varphi _{i}\left( y\right) \exp \left(
-ikx\right) /\sqrt{2\pi N}.
\end{equation*}%
\begin{figure}[tbp]
\includegraphics[width=3.5in]{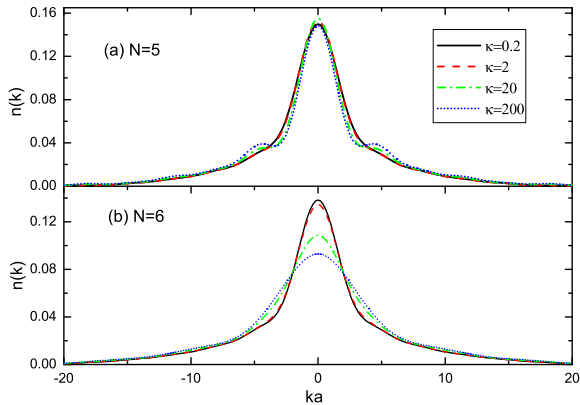}
\caption{(Color online) Momentum distribution, $n(k)$, of 5 particles (a)
and 6 particles (b) in TG limit for $\protect\kappa =0.2,2,20,200.$ When the
number of particles is odd, the secondary peaks appear for increased barrier
strength due to the interference of particle in two almost separate wells,
while for even, the momentum distribution broaden for increased barrier
strength owing to the localization of particles in separate halves of the
trap.}
\label{fig3}
\end{figure}
Fig.3 shows the momentum distribution in the TG\ limit for four different
values of the barrier strength $\kappa =0.2,2,20,200$. When $N$ is odd, as
the barrier strength is increased, the main peak firstly increases, reaches
the critical value for a barrier strength, and then begins to decrease
thereafter.\emph{\ }The secondary peaks appear with increasing strength of
barrier. The barrier separates the particles into two half wells, which,
however, are not independent but interfere with each other. It is this
interference that leads to the appearance of the secondary peaks. When the
number of particles increase, the secondary peaks become more and more
slight as a result of infinite interaction between particles. When $N$ is
even, the momentum distribution becomes broader and broader with increasing
barrier strength and the height of peak decreases monotonically, which
coincides with earlier observation that $N/2$ particles become localized in
each separate half-well and keep balance in the two sides of barrier.
\begin{figure}[tbp]
\includegraphics[width=3.5in]{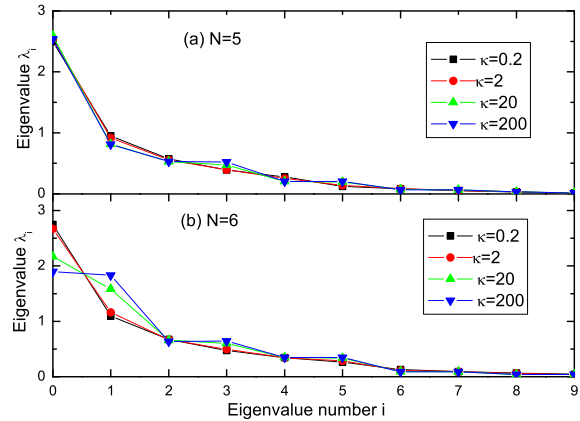}
\caption{(Color online) Occupation of the natural orbits $\protect\lambda %
_{i}$ of 5 particles (a) and 6 particles (b) in TG limit for $\protect\kappa %
=0.2,2,20,200.$ When the number of particles is odd, with the increased
barrier strength, the two largest occupations do not change almost while the
other nearby two occupations have the same value which form pairs. When $N$
is even, with the increased barrier strength, all nearby two occupations
have the same value which also form pairs.}
\label{fig4}
\end{figure}

We also present the occupation of the natural orbits for odd and even
particles in Fig.4. Clearly the presence of the split in the center will not
affect the occupation obviously for odd number of particles. The first and
second occupancies remain the same for increasing barrier height, but the
third and fourth occupancies tend to form the staircase structure, implying
the formation of pairs. Subsequently, the next two occupancies also form
pairs, and so on. Contrary to this, for even number of particles, all
occupations tend to form pairs with increasing barrier strength due to the
degeneracy of energy levels, which is also observed in fermionic TG gas \cite%
{Girardeau3}. In the case of odd number of particles, the presence of the
last particle\ prevents the formation of pairs in the the first and second
occupations.

\section{Variable Interaction Strength}

For finite particle interaction, we use exact diagonalization method \cite%
{Deuretzbacher} to deal with the many body Hamiltonian (\ref{hamltion1}),
which takes the second quantized form%
\begin{equation}
H=\sum_{i}E_{i}a_{i}^{\dagger }a_{i}+\frac{1}{2}g\sum_{ijkl}I_{ijkl}a_{i}^{%
\dagger }a_{j}^{\dagger }a_{k}a_{l}  \label{secHamilt}
\end{equation}%
where $a_{i}^{\dagger }\left( a_{i}\right) $ is the bosonic creation
(annihilation) operator for a particle in the single particle energy
eigenstate $\phi _{i}\left( x\right) $. The interaction integration
parameters $I_{ijkl}$ are calculated through $I_{ijkl}=\int dx\phi
_{i}\left( x\right) \phi _{j}\left( x\right) \phi _{k}\left( x\right) \phi
_{l}\left( x\right) $. The Hamiltonian is diagonalized in the subspace of
the energetically lowest eigenstates of non-interaction many-particle
system. Some exact results for a pair of bosons have been obtained in ref.
\cite{Murphy} and our calcualtion for even number of particles implies
similar behaviours in the relevant physical quantities. We will concentrate
on the results for odd number of bosons $N=5$ considering 26 single particle
eigenstates in the following.
\begin{figure}[tbp]
\includegraphics[width=3.5in]{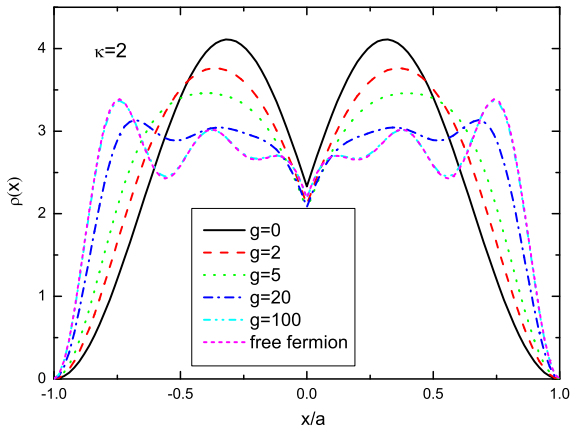}
\caption{(Color online) Particle density of five bosons for barrier strength
$\protect\kappa =2$ and different interaction strength $g=0,2,5,20,100$. The
density becomes flatter and broader with increasing $g$. In the strong
interaction regime new peaks appear gradually and finally the density tends
to be the same as that of 5 free Fermions.}
\label{fig5}
\end{figure}

Let's first discuss the particle density, $\rho \left( x\right)
=\left\langle \hat{\Psi}^{\dagger }\left( x\right) \hat{\Psi}\left( x\right)
\right\rangle =\sum_{i,j}\left\langle a_{i}^{\dagger }a_{j}\right\rangle
\phi _{i}\left( x\right) \phi _{j}\left( x\right) $ where the mean value is
calculated on the ground state of Hamiltonian (\ref{secHamilt}). Without
loss of generality, we choose the barrier strength $\kappa =2$. Fig.5 shows
the particle density for variable interaction strength $g=0,2,5,20,100$.
Because of the presence of the central barrier, the density in the middle
position appears to be lower than its nearby region. For noninteracting
bosons, there is one peak on either side of the barrier respectively. As the
interaction strength is increased, the particle density begins to spread out
and becomes flatter, while the height of two peaks is decreased. New small
peaks appear for certain interaction and finally the density tends to be the
same as that for free fermion system.
\begin{figure}[tbp]
\includegraphics[width=3.5in]{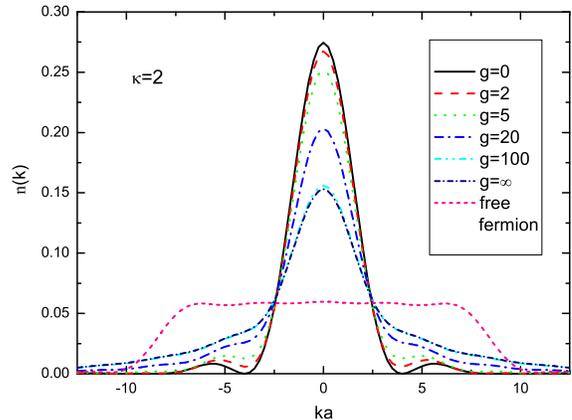}
\caption{(Color online) Momentum distribution of five bosons for different
interaction strength $g$. The secondary peaks disappear with increasing
interaction strength which hinders the interference of the particles in two
separate half-wells.}
\label{fig6}
\end{figure}

The momentum distribution for finite interaction is now calculated as $%
n(k)=\left\langle \hat{\Pi}^{\dagger }\left( k\right) \hat{\Pi}\left(
k\right) \right\rangle /N$ and is normalized to one. Here $\hat{\Pi}\left(
k\right) $ annihilates a boson with momentum $k$ and can be expanded upon
the Fourier transformation of single particle eigenstates $\phi _{i}\left(
x\right) $. Fig.6 shows the momentum distribution for the same values of
interaction strength as in Fig.5. We find, as the interaction is increased,
the secondary peaks disappear gradually and the central peak becomes lower
and broader. This result proves that the interaction will diminish the
interference between two wells. The momentum distribution is identical for $%
g=100$ and TG limit $g=\infty $ which is calculated by the Fermi-Bose
mapping in the above section. In the TG case, the long tail appears because
the higher energy levels may been occupied by some particles owing to the
infinity repel interaction, which agrees with general nature of a TG gas.
This is in strike contrast with the main feature of momentum distribution
for 5 free fermions which takes the plateau shape composed of 5 peaks and
without the long tail.
\begin{figure}[tbp]
\includegraphics[width=3.5in]{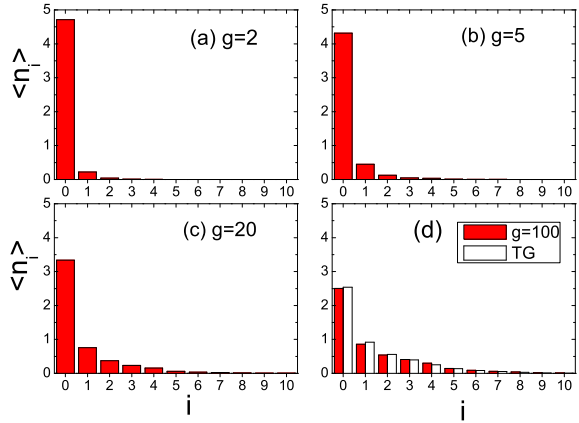}
\caption{(Color online) Occupation number distribution of five bosons for
different interaction strength $g$. On each state in (d), the left solid
part represents $g=100$ and the right hollow part shows the TG case. With
increasing $g$ the bosons leave the ground state and occupy excited states.
The occupation number in the lower state are always larger than that in the
next higher state. }
\label{fig7}
\end{figure}

We also present the result for occupation number distribution $\left\langle
n_{i}\right\rangle =\left\langle a_{i}^{+}a_{i}\right\rangle $ which is
different from the occupation of natural orbits mentioned above. The
occupied states here are single particle wavefunctions in the trap, while
natural orbits are eigenfunctions of the RSPDM. As can be seen in Fig.7,
with increasing $g$, the bosons occupy not only the ground state but also
some low-lying excited states. The occupation number in the lower state are
always larger than that in the next higher one, which is different with
particles in the harmonic trap \cite{Deuretzbacher}. In the harmonic trap,
single-particle states with even parity are stronger occupied compared to
those with odd parity. The difference between our double well and single
well structure comes from the fact that odd parity states can never feel the
existence of the split in the center. In Fig.7 (d) we compare these two
kinds of occupations of single particle states and natural orbits and find
that they almost have the same value in the corresponding state though the
details are not completely identical.

\begin{figure}[tbp]
\includegraphics[width=3.5in]{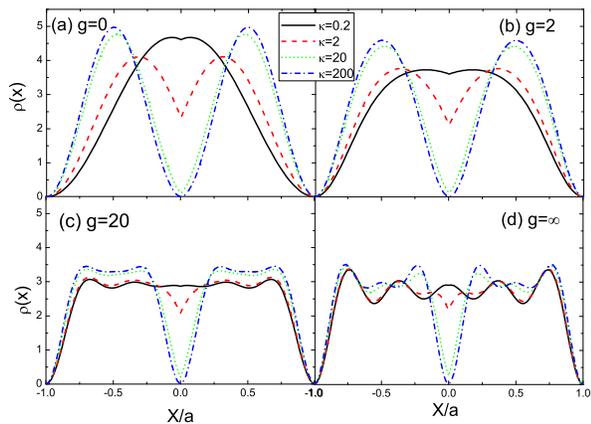}
\caption{(Color online) Particle density distribution of five bosons for
varying inter-particles interaction strength $g=0$ (a), $2$ (b), $20$ (c), $%
\infty $ (d). Particles are separated into two parts with increasing barrier
strength.}
\label{fig8}
\end{figure}

\begin{figure}[tbp]
\includegraphics[width=3.5in]{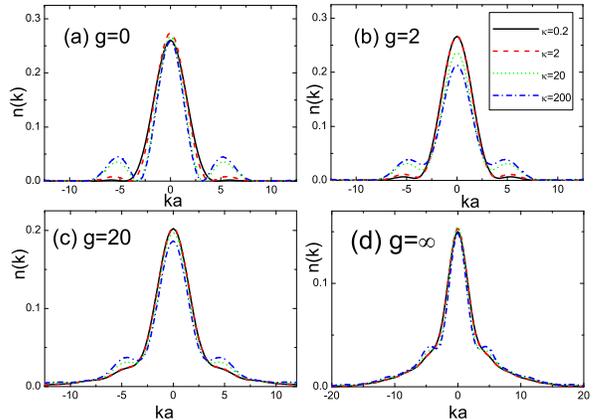}
\caption{(Color online) Momentum distribution of five bosons for varying
inter-particles interaction strength $g=0$ (a), $2$ (b), $20$ (c), $\infty $
(d). Within each plot the distribution is considered for 4 different values
of the barrier strength: $\protect\kappa =0.2$,$2$,$20$,$200$. Secondary
peaks appear with increasing barrier strength, while disappear for
increasing interaction strength.}
\label{fig9}
\end{figure}

The last knob to control the system is the variable barrier strength of the
split. Fig.8 shows the particle density distribution for four inter-particle
interaction strengths $g=0,2,20,\infty $ and four values of barrier
strengths $\kappa =0.2,2,20,200$. As the barrier strength is increased, the
density in middle area is decreasing while the density in other area is
increasing for any interaction. The barrier therefore divides the particles
into two areas regardless of the interaction. It is more constructive to say
more words on the secondary peaks in the momentum distribution, which has
already been shown in Fig.6. Physically, the interference between particles
split into the two separate half-wells gives rise to these peaks, which
reminds us the famous double-slit arrangement. Fig.9 (a) shows the momentum
distribution of 5 non-interacting particles ($g=0$) for varying $\kappa $.
In this case the momentum distribution is given by the square of the
single-particle wave-function in momentum space. So we can write it out
analytically%
\begin{eqnarray}
n\left( p\right) &=&C(\frac{\sin \left( k+p\right) }{k+p}+\frac{\sin \left(
k-p\right) }{k-p}+\frac{\kappa }{k}\frac{1-\cos \left( k+p\right) }{k+p}
\notag \\
&&+\frac{\kappa }{k}\frac{1-\cos \left( k-p\right) }{k-p})^{2}
\end{eqnarray}%
with $C$ is a normalization factor. The secondary peaks can be seen more and
more clearly with increasing strength of barrier. Strength of the barrier
causes the spreading of the particle-density in two wells, which in turn
leads to a narrowing of momentum peaks. For finite interaction strength,
Fig.9 (b-d) show similar tendency for the central and secondary peaks. The
disappearance of the secondary peaks with increasing interaction strength
can be attributed to the increased localization of the particles, analogous
to the superfluid-insulator transition of the Bose-Hubbard model in optical
lattice studies \cite{Greiner}.

\section{Conclusion}

In conclusion, we have studied the ground state property of few bosons in a
hard wall trap with $\delta -$type split in the center. Based on the
Fermi-Bose mapping, the reduced single-particle density and the momentum
distribution of odd $(N=5)$ and even $(N=6)$ bosons are illustrated and
explained in TG limit. The off-diagonal quadrants representing the
interference between two half wells remain nonzero for $N$ is odd, while
disappear for $N$ is even. For finite interaction strength, the exact
diagonalization method enables us to manipulate the system in several ways,
including variable number of particles, interaction strength, barrier height
of the split, etc. The interference between particles on the two sides of
the split manifests itself in the secondary peaks of the momentum
distribution, emerging with increasing barrier strength $\kappa $, while
vanishing for increasing interaction strength $g$.

\begin{acknowledgments}
This work is supported by NSF of China under Grant No. 10774095 and
10574150, 973 Program under Grant No. 2006CB921102, ``Bairen'' Programs of
Chinese Academy of Sciences, and Shanxi Province Youth Science Foundation
under Grant No. 20051001.
\end{acknowledgments}

\end{document}